\documentstyle[12pt,epsfig]{article}

\begin{document}
\title{Vacuum Polarization in the Spacetime of a 
Scalar-Tensor Cosmic String}
\author{V. B. Bezerra$^{1}$\thanks{valdir@fisica.ufpb.br}, 
R. M. Teixeira Filho$^{1,2}$\thanks{rmuniz@fisica.ufpb.br} \\ 
G. Grebot$^{3}$\thanks{guy@mat.unb.br} and  
M. E. X. Guimar\~aes$^3$\thanks{emilia@mat.unb.br} \\
\mbox{\small{1. Departamento de F\'{\i}sica, UFPb, 58051-970, 
Jo\~ao Pessoa, Pb, Brazil}} \\
\mbox{\small{2. Instituto de F\'{\i}sica, UFBA, 40210-340, 
Salvador, BA, Brazil}} \\
\mbox{\small{3. Departamento de Matem\'atica, UnB, 70910-900, 
Bras\'{\i}lia, DF, Brazil}}}
\date{}
\maketitle
\begin{abstract}
We study the vacuum polarization effect in the spacetime 
generated by a magnetic flux cosmic string in the framework 
of a scalar-tensor gravity. The vacuum expectation values 
of the energy-momentum tensor of a conformally coupled scalar 
field are calculated. The dilaton's contribution to the vacuum 
polarization effect is shown explicitly.
\end{abstract}
\section{Introduction}
One of the most interesting features of the spacetime generated 
by a static, straight axially symmetric cosmic string in General 
Relativity \cite{li}
\begin{equation}
ds^2 = - dt^2 + dz^2 + d\rho^2 + B^2 \rho^2 d\varphi^2  
\end{equation}
with $B =1- 4\mu G $, is that particles and fields are sensitive 
to its global (conical) structure and, therefore, some physical 
effects may arise due solely to 
the conicity of this geometry. In particular, many 
authors \cite{many} have already considered the vacuum polarization 
effect in connection with the Casimir effect \cite{bir} in which 
the conducting planes form an angle equal to the deficit angle 
$\Delta = 8\pi\mu G$ associated with metric (1). In the papers 
\cite{frolov,meli} a more general situation has been carried out. Namely, 
a cosmic string carrying a magnetic flux interacting with a 
charged scalar field placed in background (1) was considered. In 
this case, the vacuum polarization effect arises not only in 
connection with the non-trivial gravitational interaction but also 
with the Aharonov-Bohm interaction. 

It is interesting to notice that all the above mentioned 
implications of the interactions between a quantum field and a 
cosmic string have been done in the framework of Einstein's gravity. 
However, it has been argued that gravity may not be described by a 
purely tensorial field $g_{\mu\nu}$. In particular, the existence 
of a scalar partner $\phi$ (for instance, the dilaton field) 
for the graviton arises naturally in all attempts to unifying 
gravity with the other fundamental interactions \cite{green}. 
Although Einstein's theory agrees with its experimental tests with an 
accuracy of one percent or better, this present agreement 
between theory and experiments is compatible 
with the existence of a long-range scalar (gravitational) field: 
it has been shown that General Relativity acts as an attractor 
to the scalar-tensor gravities as a consequence of the cosmological 
expansion which drives the scalar couplings towards zero \cite{dam,dam2}. 

In this letter, we are interested in studying the vacuum polarization effect 
of a charged, (massless) scalar field due to a magnetic flux 
cosmic string in the framework of the scalar-tensor theories of 
gravity. For this purpose, we will first present the metric 
generated by a cosmic string in scalar-tensor gravities \cite{guima} 
and we will show that this metric is conformally flat,  to linear order 
of $G_0\mu$. Then, we compute the vacuum expectation values (v.e.v.) 
of the components of the energy-momentum tensor of a 
conformal scalar field. Whenever convenient, we reduce our 
results to the particular case of the Brans-Dicke gravity and we 
compare these results with the ones obtained in the 
framework of Einstein's gravity. We anticipate that our 
main result is to derive explicitly the dilaton's contribution to 
the vacuum polarization effect. 

This work is outlined as follows. In section 2  we recall 
some results which will be used throughout this paper  compute 
the v.e.v. of the stress-energy tensor for a conformally coupled scalar 
field. We reduce our results to the particular case of the 
Brans-Dicke theory of gravity and we present graphs of 
the energy  density for a neutral ($\gamma =0$) and a 
twisted ($\gamma = 1/2$)  conformal scalar fields in the 
Brans-Dicke theory and we show that the Aharonov-Bohm interaction 
is the leading interaction between the scalar field and 
the magnetic flux cosmic string. Finally, in  section 3 we 
end with some conclusions and discussions on the results of the present work.

\section{The Vacuum Polarization Effect in Scalar-Tensor Gravities}

The metric of a static, straight axially symmetric cosmic string 
in scalar-tensor gravity is \cite{guima}
\begin{equation}
ds^2 = \left[ 1+ 8G_0\mu \alpha^2(\phi_0)\ln \rho/\rho_c \right] 
\left[ -dt^2 + dz^2 + d\rho^2 + (1- 8G_0\mu)\rho^2 d\varphi^2 \right] ,
\end{equation}
where $G_0$ is defined as $G_0 \equiv G_*A^2(\phi_0)$ and 
$\alpha(\phi)= \partial \ln A/\partial\phi$ is the coupling 
between matter and the dilaton field. All quantities here are 
computed up to first order in $G_0\mu$. The constant $\rho_c$ 
appearing in metric (2) is a constant of integration and is, 
conveniently, of the same order of magnitude of the string's radius. 
$\phi_0$ denotes the cosmologically-determined value of the 
dilaton field far away from the solar system.  

Let us define the conformal factor
\[
\Omega \equiv 1 + 4G_0\mu \alpha^2(\phi_0)\ln \rho/\rho_c , 
\]
and denoting $B = 1 - 4G_0 \mu$, metric (2) can be re-written as 
\begin{equation}
ds^2 = \Omega^2_{lin} [ -dt^2 + dz^2 + d\rho^2 + B^2_{lin}\rho^2 d\varphi^2] .
\end{equation}
$\Omega^2_{lin}$ is 
the linearised conformal factor; its expression being 
$\Omega^2_{lin} = 1+ 8G_0\mu \alpha^2(\phi_0)\ln \rho/\rho_c$ 
and $B^2_{lin}$ is given by $B^2_{lin} = 1 - 8G_0\mu$. Let $\theta$ be 
the new azimuthal angle 
$\theta = (1-4G_0\mu)\varphi$. Then, metric (3) becomes 
conformally flat with deficit angle equal to $\Delta \theta = 8\pi\mu G_0$. 

Since metric (3) is conformally flat, we can apply an 
alternative expression to compute the v.e.v. of the components 
of the energy-momentum tensor  $<T^{\mu}_{\nu}>$, instead of making 
use of the Green's functions \cite{bir}. Namely, in  the particular 
case of a conformally coupled scalar field ($\xi = 1/6$ in 4-dimensions), 
we have
\begin{equation}
<T^{\mu }_{\nu }>_{\bar{g}} =(\frac{g}{\bar{g}}%
)^{1/2}<T^{\mu }_{\nu }>_{g}-\frac{1}{2880\pi ^{2}}[%
\frac{1}{6}\,^{(1)}H_{\mu }^{\nu }-\,^{(3)}H_{\mu }^{\nu }] , 
\end{equation}
where
\[
^{^{(1)}}H_{\mu\nu }\equiv 2R_{;\mu \nu }-2\bar{g}_{\mu \nu
}\Box_{\bar{g}} R-\frac{1}{2}\bar{g}_{\mu \nu }R^{2}+2RR_{\mu \nu }
\]
\[
^{\,^{(3)}}H_{\mu\nu }\equiv \frac{1}{12}R^{2}\bar{g}_{\mu
\nu }-R^{\rho \sigma }R_{\rho \mu \sigma \nu }.
\]
For the seek of clarity, we have denoted metric (3) as 
$\bar{g}_{\mu\nu}$ in order to distinguish from the metric 
$g_{\mu\nu}$ of the flat spacetime (1). The term 
$<T_{\mu }^{\nu }>_{g}$ appearing in the r.h.s. of 
expression (4) is the energy-momentum tensor computed with 
respect to metric (1) and has been already calculated in the ref. \cite{meli} 
\[
<T^{\mu }_{\nu }>_{g} = \left[ \omega_4(\gamma) - \frac{1}{3} 
\omega_2(\gamma)\right] \frac{1}{\rho^4} diag(1,1,1,-3) \, ,
\] 
The quantities $\omega_2(\gamma)$ and $\omega_4(\gamma)$ 
were evaluated by Dowker  \cite{frolov,dow}
\[
\omega_2(\gamma) = -\frac{1}{8\pi^2} \left\{ \frac{1}{3} -\frac{1}{2B^2}
\left[ 4\left(\gamma -\frac{1}{2}\right)^2 -\frac{1}{3} \right] \right\} ,
\]
\begin{eqnarray*}
\omega_4(\gamma) & = & -\frac{1}{720\pi^2}\{ 11 -\frac{15}{B^2} 
\left[ 4\left( \gamma -\frac{1}{2} \right)^2 -\frac{1}{3} \right] \nonumber \\
& & + \frac{15}{8B^4} \left[ 16 \left( \gamma - \frac{1}{2}\right)^4 
- 8\left( \gamma - \frac{1}{2}\right)^2 +\frac{7}{15} \right] \} .
\end{eqnarray*}
Both expressions are valid only if $B>1/2$. $\gamma$ is the 
fractional part of $\Phi/\Phi_0$, $\Phi_0$ being the quantum 
flux $2\pi/e$, and lies in the interval $0\leq \gamma < 1$. The 
particular values of $\gamma =0$ and $\gamma =1/2$ correspond to 
the cases of a vanishing flux and a twisted field around the 
axis $\rho=0$, respectively. 

Therefore, for a conformally coupled scalar field in the 
spacetime (3), we have\footnote{Expression (5) was obtained with 
the help of the computer algebra program {\it Maple}.}, up to 
second order in $G_0\mu$: 
\begin{eqnarray}
<T^{\mu }_{\nu }>_{\bar{g}} & =&  \left(1- 16G_0\mu\alpha^2(\phi_0)
\ln\rho/\rho_c+ 128{G_0}^2\mu^2\alpha^4(\phi_0)\ln^2\rho/\rho_c  \right)
<T^{\mu }_{\nu }>_{g} \nonumber \\
& & - \frac{1}{15\pi^2\rho^4}G_0^2\mu^2\alpha^4(\phi_0) 
diag (\frac{1}{3},\frac{1}{3},-\frac{1}{3},1).
\end{eqnarray}

The vacuum polarization effect expressed by (5) is a consequence 
of the conical geometry (non-trivial gravitational interaction), 
of the Aharonov-Bohm interaction between the quantum scalar field 
and the magnetic flux string, and of the presence of the dilaton field in
this theory. Expression (5) is convenient because it expresses 
the contribution of the dilaton field to the vacuum polarization 
effect explicitly. The second term in the r.h.s. of (5) is a contribution 
due solely to the dilaton in comparison to the first term which 
is a combination of all the interactions. It is interesting 
to notice that such a contribution is fully described by 
one dimensional coupling strength $(G_{0})$ and one post-Newtonian 
parameter $(\alpha (\phi_0))$. Finally, we point out that the 
trace anomaly appears up to second order in $G_0\mu$. 

\subsection*{The Particular Case of the Brans-Dicke Theory}

It is very illustrative to  consider a particular form for the 
coupling function $\alpha(\phi)$, corresponding to the Brans-Dicke 
theory. Namely, $\alpha^{2} = \frac{1}{2 \omega + 3}, (\omega = cte)$. 
In this case, the metric of a cosmic string is given, to first 
order in $G_0\mu$, by \cite{romero}:
\[
ds^2 =   \left[ 1 + \frac{8\mu G_{0}}{2\omega + 3}
\ln\frac{\rho}{\rho_c} \right] [-dt^2 + dz^2 + d\rho^2 + 
(1 - 8\mu G_0) \rho^2 d\theta^2 ] .
\]
Besides, we have that  
$G_{0}= \left( \frac{2 \omega + 3}{2 \omega + 4} \right) G$ 
where $G$ is the Newtonian constant \cite{bd}. Therefore,  
expression (5) reduces to:
\begin{eqnarray}
<T^{\mu }_{\nu }>_{\bar{g}} & = & \left[1- \frac{16G\mu}{2\omega+4}
\ln\rho/\rho_c + \frac{128G^2\mu^2}{(2\omega+4)^2}\ln^2\rho/\rho_c\right]
\left[ \omega_4(\gamma) - \frac{1}{3} \omega_2(\gamma)\right] 
\frac{1}{\rho^4} diag(1,1,1,-3) \nonumber \\
& & - \frac{1}{15\pi^2\rho^4}\frac{G^2\mu^2}{(2\omega + 4)^2}
diag (\frac{1}{3}, \frac{1}{3},-\frac{1}{3},1) .
\end{eqnarray}
We can  verify that in the limit where $\omega \rightarrow \infty$ 
our result agrees with the one obtained in the framework of General 
Relativity, as expected. 
For values of $\omega$ such that $\omega > 2500$ (consistent with 
solar system experiments made by Very Long Baseline  Interferometry 
(VLBI) \cite{eub}), we can  see that the corrections due to the 
presence of the dilaton are very small in comparison with the 
previous situation in General Relativity. 

Figures 1 and 2 present the behaviour of  the energy density of a 
massless, conformally coupled ($\xi=1/6$) scalar field in the 
particular cases of a vanishing flux ($\gamma =0$) and a 
twisted field ($\gamma = 1/2$) in the Brans-Dicke gravity, 
respectively. We can notice that the Aharonov-Bohm interaction is 
the leading interaction between the scalar field and the magnetic 
flux cosmic string. 

\section{Conclusions}

In this work, we have computed the vacuum expectation values of 
the energy-momentum tensor of a conformally coupled scalar field, by 
noting that spacetime (3) is conformally flat. The expression for this 
tensor (5) reveals explicitly the dilaton's contribution to the 
vacuum polarization effect. As an example, we considered the particular 
case of the Brans-Dicke theory and we presented the behaviour of the 
energy density for a conformally coupled scalar field for both 
$\gamma=0$ and $\gamma =1/2$ cases. The Aharonov-Bohm interaction is 
the predominant interaction between the (charged) scalar field and the 
magnetic flux cosmic string, a result which is also valid in the 
framework of the  General Relativity theory \cite{me}.

\section*{Acknowledgments}
MEXG thanks Dr. Luiz Paulo Colatto for helpful discussions and a 
critical reading of this manuscript. VBB thanks CNPq for partial 
financial support.

\begin{figure}[h]
\label{fig1}
%\vspace{5cm}
%\hspace{3cm}
%\seteps{0cm}{7cm}{5cm}{GRAFIC~8.eps}
%\special{eps:/home/guy/letter/graficos08.eps x=7cm y=5cm}
%\includegraphics[width=7cm]{graficos08.eps}
\includegraphics{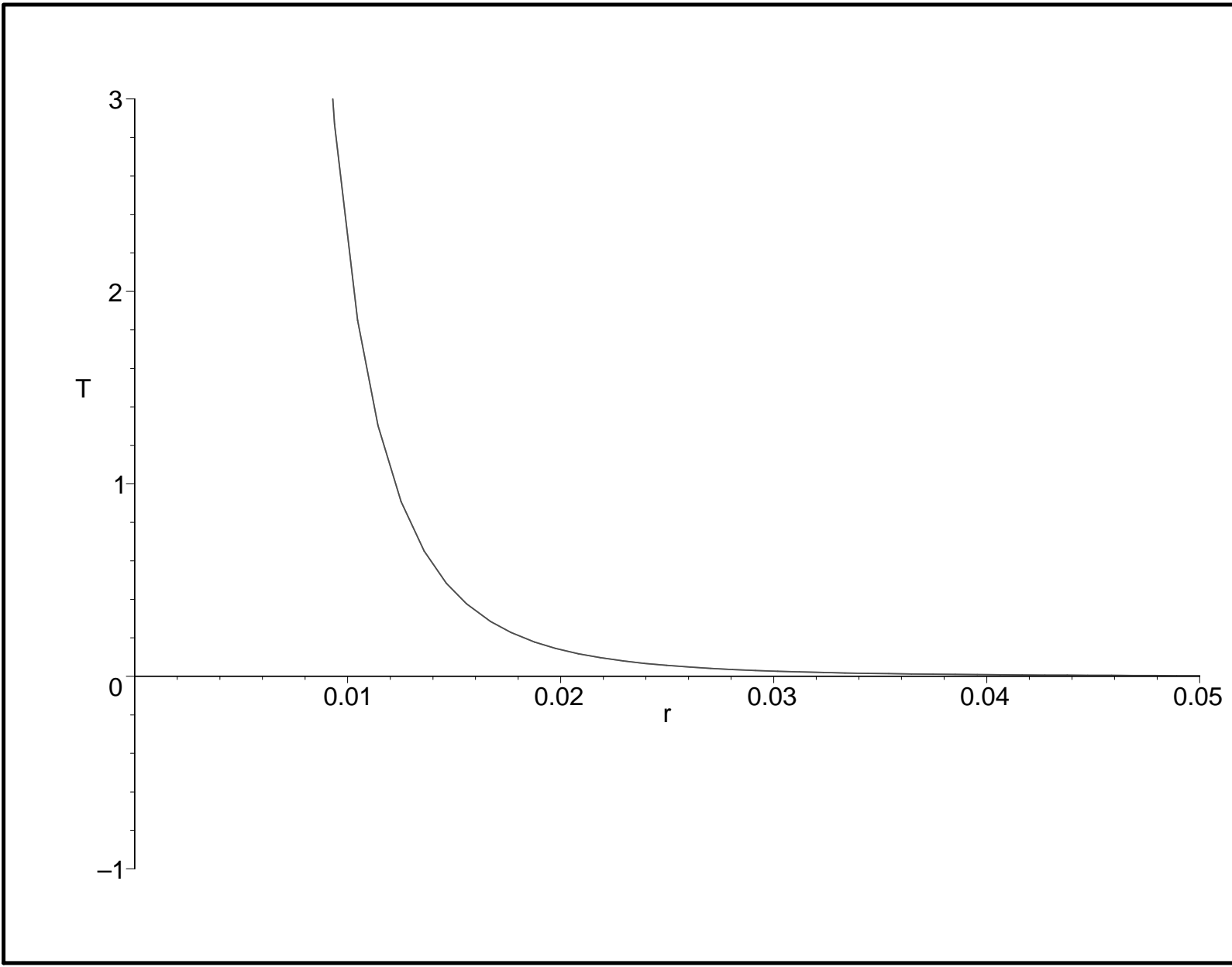}
\caption{Graph of $\pi^2\,<T_{00}>$, with $\gamma=0$ and $\omega=3000$}
\end{figure}
\begin{figure}[h]
\label{fig2}
%\vspace{5cm}
%\hspace{3cm}
%\seteps{0cm}{7cm}{5cm}{GRAFIC~9.eps}
\includegraphics{graficos09.eps}
%\special{eps:/home/guy/letter/graficos09.eps x=7cm y=5cm}
\caption{Graph of $\pi^2\,<T_{00}>$, with $\gamma=1/2$ and $\omega=3000$}
\end{figure}

\end{document}